\documentclass[dvips]{article}
\usepackage{graphicx}
\usepackage{amssymb}
\def\vsigma{\vec\sigma}

\def\vtau{\vec\tau}
\def\vnu{\vec\nu}

\def\<{\langle}
\def\>{\rangle}

\def\det{{\rm det}}

\def\rv{{\bf r}}

\def\Rv{{\bf R}}

\def\vv{{\bf v}}

\def\Bv{{\bf B}}
\def\Ev{{\bf E}}

\def\zu{\hat{\bf z}}
\def\lambdabar{\lambda\raise0.4ex\hbox{\kern-0.5em\hbox{--}}\ }

\def\Fv{{\bf F}}

\def\F{{\cal F}}
\def\E{{\cal E}}
\def\B{{\cal B}}

\def\rot{\nabla\times}
\def\div{\nabla\cdot}

\def\ni{\noindent}
\def\be{\begin{equation}}
\def\ee{\end{equation}}

\begin{document}

\bigskip
\centerline{PARTICLE ACCELERATION IN A HELICAL WAVE GUIDE} 
\vskip 0.7 true cm

\centerline{X. Artru, C. Ray}

\vskip 0.5 true cm

\centerline{Institut de Physique Nucl\'eaire de Lyon, Universit\'e de Lyon,} 

\centerline{Universit\'e Lyon 1 and CNRS/IN2P3, F-69622 Villeurbanne, France}

\centerline{e-mails: x.artru@ipnl.in2p3.fr ; c.ray@ipnl.in2p3.fr}

\vskip 0.6 true cm \noindent
\textbf{Abstract.}
The electromagnetic wave field propagating in a helical wave guide is decomposed in an angular momentum basis. Eigenmodes are calculated using a truncation in $l$ and a discretisation of the boundary condition. Modes slightly slower than light are considered for relativistic particle acceleration. 

\vskip 0.6 true cm \noindent

\section{Introduction}

An electromagnetic wave can accelerate charged particles over long enough distance if 1) it has a longitudinal electric component 2) its phase velocity is equal or close to the particle velocity, therefore less than $c$. 
A free plane wave misses the first condition and a wave guided by a metallic straight tube miss the second condition. The usual solution to satisfy both conditions is a chain of coupled resonating cavities. Another solution is the helically bent tube. Here the "photon path" is longer than the corresponding path in a straight tube, so that some modes can be made slower than light, while having a non-zero longitudinal electric field on the helicity axis.

A helical wave guide is defined by its axis $Oz$, its period $\Lambda=2\pi/q$ and its \emph{base curve} $\rv(s)=(X(s),Y(s))$ which is the intersection of the wall with the $z=0$ plane. The wall surface is generated by helices $H_s$ of parametric equation
\be\label{param}
(x,y,z)=({\cal R}(qz)\rv(s),z)
\ee 
where the $2\times2$ matrix ${\cal R}(qz)$ describes the rotation of angle $qz$ about $0z$.

The traveling wave tube, used for generation of radar waves, is a special kind of helical wave guide. It is made of a cylindrical metallic tube inside which a thick conducting helical wire is fixed. Its base is not simply connected~: it is a disk with a hole corresponding to the section of the wire at $z=0$. Therefore the base curve has two disconnected pieces~: the circular base of the cylinder and, inside it, a small closed curve. The same device, with a superconducting helix, has been studied for particle acceleration \cite{HELIX,Aron:1973ya}.

A tube where the wall itself possesses a helical structure can be made by drawing helical grooves on the internal face of an initially cylindral tube. In the case where the groove section is made of rectangular steps, the propagation modes can be calculated using an approximate field-matching technique \cite{Wei}, or a 3-dimensional computer code \cite{MAFIA}.

\begin{figure}\label{courbase} 
\centering{%
\includegraphics[width=.2\textwidth,bb= 300 150 500 500]{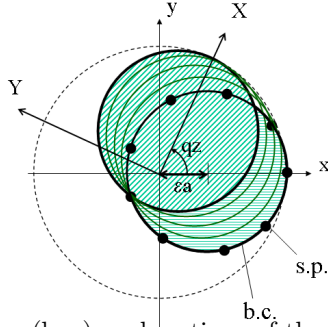} 
}
\caption{Base curve (b.c.) and sections of the helical wave guide at successive values of $z$. The rotating frame $\{OX,OY\}$ of the last section is shown. In this example the base curve is an ex-centered circle of radius $a$. The balls (s.p.) represent the sutture points.}
\end{figure}

In this paper we will consider helical tubes which have smoothly curved walls, associated to smooth base curves. To calculate the modes we will use a truncated partial wave expansion and enforce the boundary conditions only at discrete points of the base curve, called \emph{sutture points},  the number of sutture points being equal to the number of partial waves. For definiteness, we will take the base cuve to be a ex-centred circle, as shown in Fig.\ref{courbase}. The guide parametres are then
\begin{itemize}
	\item $a$ = radius of the base circle,
	\item $\varepsilon\in[0,1]$ = relative amplitude of the helical deformation. The centre of the base circle is at distance $\varepsilon a$ from $Oz$.
        \item $q$ = $2\pi$/(helix period).
\end{itemize}
The average angle of an helix with $Oz$ (\emph{i.e.,} ${\pi\over2}$ minus the pitch angle) is $\bar\alpha= \arctan(qa)$. Another significant quantity is the maximum helix angle of the wall with $Oz$, $\beta=\arctan(\varepsilon qa)$. The case $\beta=0\ $ ($q=0$ or $\varepsilon=0$) corresponds to a straight cylindrical guide.

\subsection{The scalar field case}
We first study the simpler case of a scalar field $\Phi(t,x,y,z)$, which obeys the Klein-Gordon equation (in units $c=1$) 
\be
(-\partial^2_t+ \partial^2_x+ \partial^2_y+ \partial^2_z)\, \Phi(t,x,y,z)=0\,,
\ee
with the boundary condition
\be
\Phi(t,x,y,z)=0 \quad\hbox{on the wall} \,.
\ee
In a fixed $z$ transverse plane we introduce a \emph{"rotating frame"} whose basis vectors rotate about $Oz$ by the angle $qz$ when $z$ increases. The fixed- and rotating-frame coordinates, $\rv=(x,y)$ and $\Rv=(X,Y)$, are related by $x+iy = e^{iqz}\,(X+iY)$. The corresponding polar coordinates are $(r,\varphi)$ and $(r,\varphi-qz)$ respectively. We look for a solution of the form
\be\label{reduction-sca}
\Phi(t,x,y,z)=e^{iPz-i\omega t}\, \Psi(X,Y) \,.
\ee
$P$ is the longitudinal pseudo-momentum (in the Bloch wave sense).  It is the eigenvalue of the operator 
\be\label{Phelic}
{\cal P}_{\rm helic}=p_z+q{\cal J}_z=-i\partial/\partial_z-iq\partial/\partial_\varphi
\ee
which is conserved due to the invariance under the helical displacement

\be \label{transf-helic}
\{\hbox {translation by }\Delta z\} \times 
\{\hbox {rotation by } q\Delta z\} \,.
\ee 

\ni
For a given $\omega$ the Klein-Gordon equation with the boundary condition has a finite number of solutions or \emph{modes}. $P$ and $\omega$ are linked by the dispersion relation $P=P_m(\omega)$ or $\omega=\omega_m(P)$. The phase and group velocities are $v_{\rm ph} = \omega/P$, $\ v_{\rm gr} = d\omega/dP$.

Since  $\Phi$ is an eigenvector of (\ref{Phelic}) we can decompose it in free cylindrical  partial waves of angular momenum $l$ and longitudinal momentum $p_l=P-lq$ 
\be\label{decompose-scalaire}
\Phi(t,x,y,z)=e^{-i\omega t}\,\sum_{l=-\infty}^{+\infty}   a_l \, e^{il\varphi}\, e^{i(P-lq) z} \,\psi_{l,k(l)}(r) \,,
\ee
where
\be
k(l)\ {\rm or}\ k_l=(\omega^2-p_l^2)^{1/2} =\left[\omega^2-(P-lq)^2\right]^{1/2}
\ee
is the transverse momentum of the $l$ component and $\psi_{l,k}(r)$ is a solution of the radial equation
\be\label{eq-radiale}
\psi''+\psi'/r+(k^2-l^2/r^2)\psi=0\,.
\ee
$k_l$ is real for 
\be\label{kl=real}
{(P-\omega)}/q \le l \le{(P+\omega)}/q\,,
\ee
otherwise it is pure imaginary : $k_l=i\kappa_l$, with $\kappa_l=\left[(P-lq)^2-\omega^2\right]^{1/2}$. 

The solution of (\ref{eq-radiale}) is proportional to the Bessel  function $J_l(kr)$ or the modified Bessel function $I_l(\kappa r)\equiv J_l(i\kappa r)/i^l$. We will fix the normalisation of $\psi_{l,k}(r) $ by
\be\label{Bessel}
\psi_{l,k}(r) = J_{|l|}(kr)/k^{|l|} = I_{|l|}(\kappa r)/\kappa^{|l|} \,.
\ee
With $k=i\kappa$ the last two expressions are equivalent. The $k^{|l|}$ denominator allows a finite limit for $\psi_{l,k}(r) $ when $k\to0$.   

Let us write the boundary condition for $t=z=0$. Il will be satisfied also at all $t$  due to the invariance under time translation and all $z$ due to the invariance under helical displacement. For every point $\rv(s)=(X(s),Y(s))$  of the base curve one has the constraint
\be
\sum a_l\, e^{il\varphi(s)}  \, J_{|l|}[kr(s)]/k^{|l|}=0\,,
\ee
where $(r(s),\varphi(s))$ are the polar coordinates of $\rv(s)$.

\subsection{The truncation-and-sutture method} 
One keeps only a finite set $\mathcal{L}$ of partial waves and selects an equal number of "sutture points" $\rv_1,...\rv_N$ of the base curve (see Fig.\ref{courbase}). For instance $\mathcal{L}=[l_{\rm inf},l_{\rm sup}]$ with $l_{\rm sup}=l_{\rm inf}+N-1$. The $a_l$ are approximatively determined by the system of linear equations 
\be\label{eq-lin-scal}
\sum_{l\in\mathcal{L}} a_l\, e^{il\varphi_n} \, J_{|l|}(kr_n)/k^{|l|}=0\,,  
\quad 1 \le n \le  N \,,
\ee
$\ (r_n,\varphi_n)$ being the polar coodinates of $\rv_n$. A nonzero solution exists if the determinant est zero:
\be\label{det-scal}
\det\{C_{nl}\}=0 \,,\ {\rm where}\ \ C_{nl}= e^{il\varphi_n} \, J_{|l|}(kr_n)/k^{|l|} \,.
\ee
This is the dispersion relation in the truncation-and-sutture approximation. 
The following conditions may be imposed on $\mathcal{L}$ and the $\rv_n$'s ~:
\begin{itemize}
	\item the $\rv_n$'s should give a precise enough definition of the base curve,
	\item $l \, (\varphi_{n+1}-\varphi_n)$  should not be too large compared to unity,
	\item $k_l \, (r_{n+1}-r_n)$ should not be too large compared to unity.
\end{itemize} 
The ultimate criterium of the quality of the approximation is the stability of the roots of (\ref{det-scal}) when $N$ is changed by one or two units. The coefficients $a_l$ should be stable also, at least for $l$ far enough from the border of $\mathcal{L}$.  

\paragraph{Case of a symmetric base.}
If the base is symmetrical about $Oz$, the modes are either even or odd under this symmetry.  One can study the even and odd modes separately by taking either even or odd $l$'s in $\mathcal{L}$. Similarly, if the base has a ternary symmetry (invariance under rotation by $2\pi/3$), $\mathcal{L}$ can be restricted to either $l=0$ or $l=1$ or $l=2$ modulo 3. To avoid redundant equations in (\ref{det-scal}), the points $\vec r_n$ should be all different modulo the symmetry. One can choose them on one part of the base curve only, the other parts being deduced by the symmetry. 

\subsection{Numerical example in the scalar case}\label{numeric-scal}
 We took a circular base with parameters $\varepsilon=0.4$, $q=1$ and imposed the phase velocity $v_{\rm ph} = \omega/P=1-10^{-5}$. Plotting $\ln|\det\{C_{nl}\}|$ as a function of $\omega a$ for different sets $\mathcal{L}$ and equidistant $\rv_n$'s, with $N$ ranging from 5 to 21, we found zeros at $\omega a=5.0$, $7.0$ which are stable at the $0.1$ precision when $\mathcal{L}$ contains at least the set $[-1,+5]$. Another root, $\omega a=8.5$, seems to stabilise for $\mathcal{L}\supset[-1,+7]$. Higher modes are seen, but would require more partial waves to get stable. 
Fig.2 shows the scalar field $\Psi(X,Y)=\Phi(0,X,Y,0)$ of the $\omega a=5.0$ mode. Due to the low number of sutture points, the boundary condition $\Psi(X,Y)=0$ at the boundary is only appoximately satisfied. 
One observes that $\Psi(X,-Y)=\Psi^*(X,Y)$. This property comes from the symmetry of the base about the $X$ axis. 
The maximum of $|\Re\Psi|$ is not at the centre of the base, but farther from the helix axis. This is a manifestation of the centrifugal effect. One can also observe that the phase $\arg\Psi=\arctan(\Im\Psi/\Re\Psi)$ increases with the azimuth, which means that the average angular momentum is positive, as expected.

\begin{figure}
\centering{%
\begin{tabular}{c c}
\includegraphics[width=.49\textwidth]{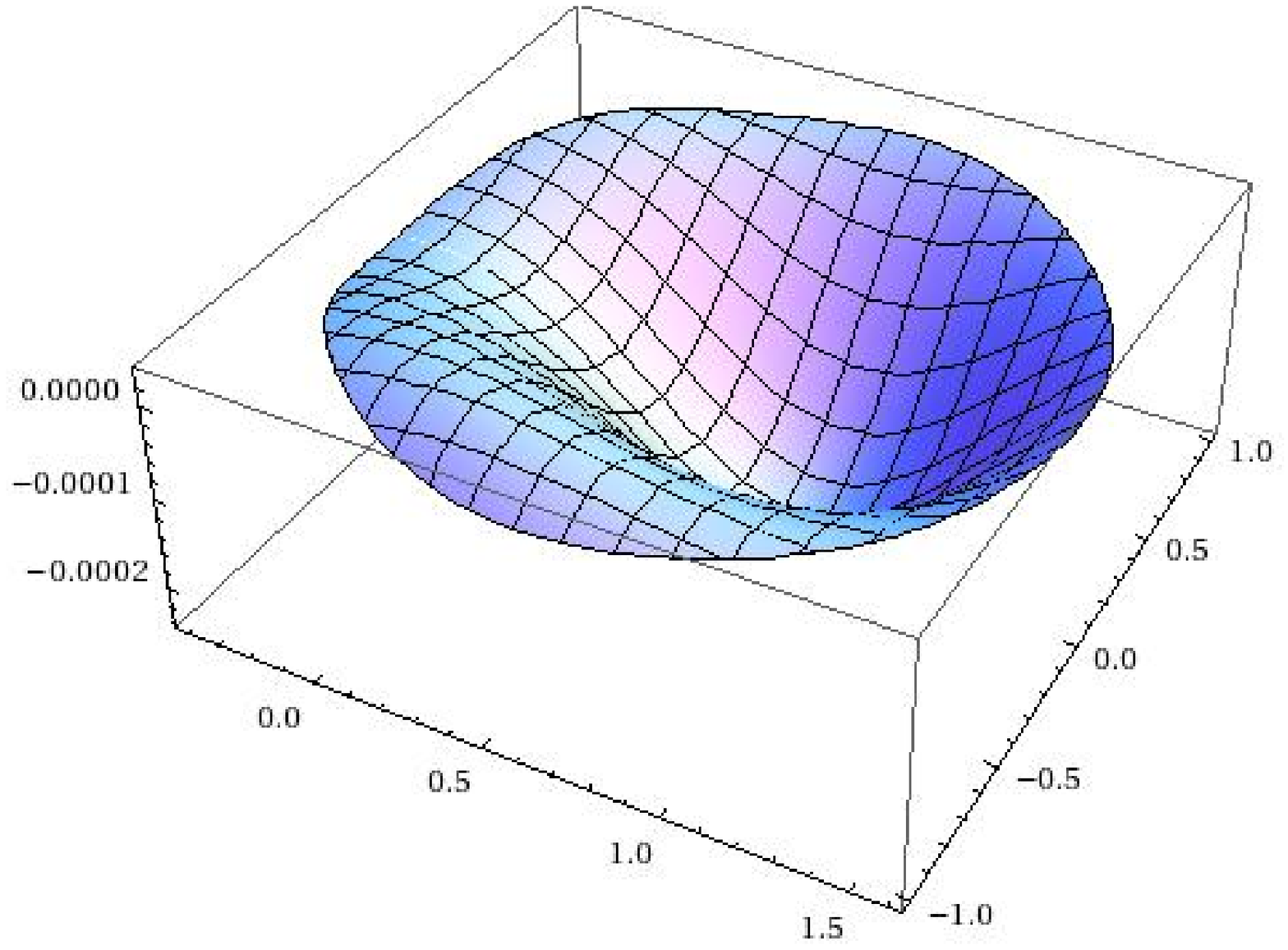} 
&
\includegraphics[width=.49\textwidth]{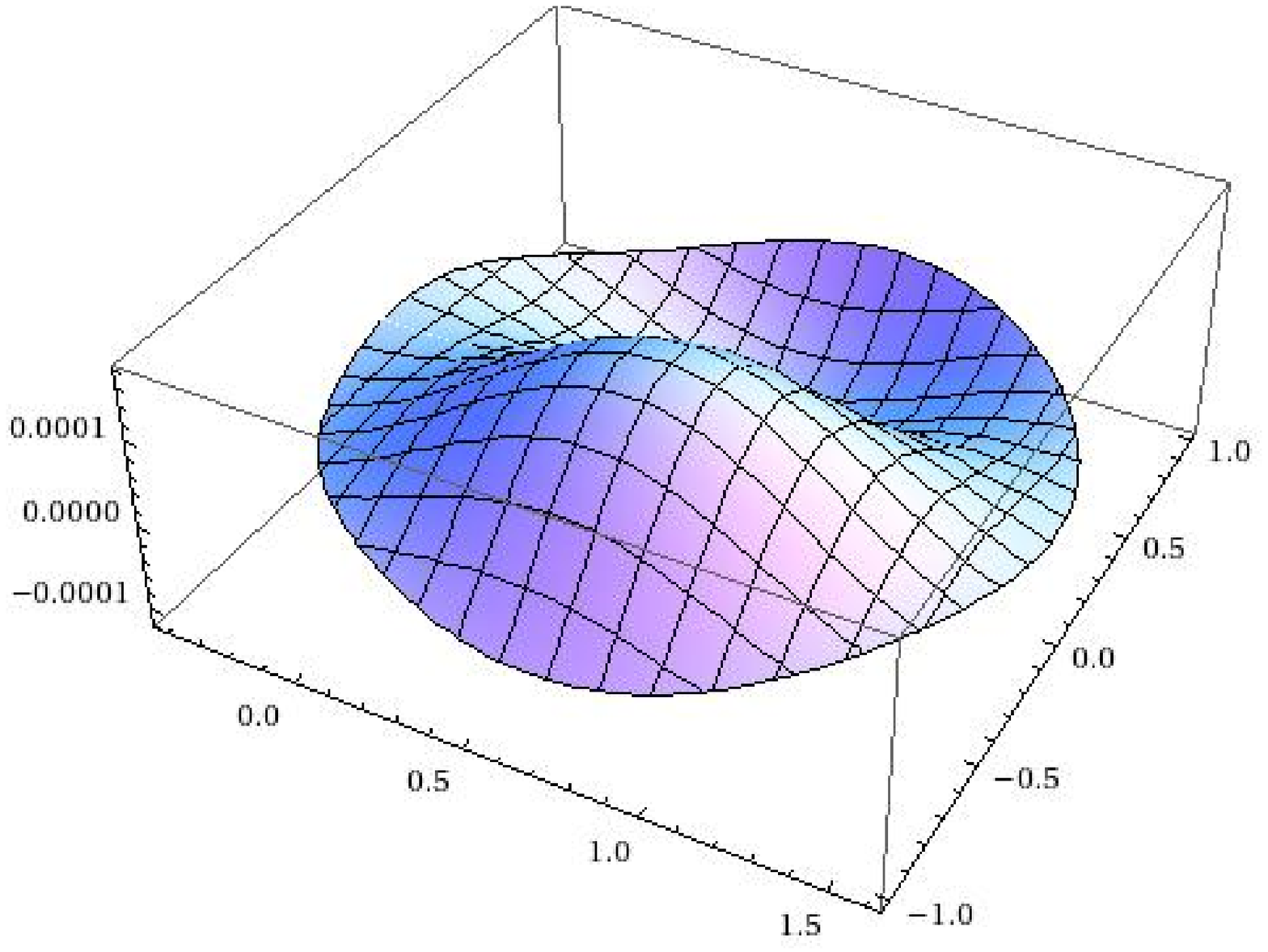} 
\end{tabular}
}
\caption{3-dimensional representation of the scalar field $\Psi(x,y)=\Phi(0,x,y,0)$ of the $\omega a=5.0$ mode, for a circular base with $\varepsilon=0.4$ and $q=1$. 
Left: real part; right: imaginary part.}
\end{figure}

\section{The Maxwell field}

We replace the Klein-Gordon equation by the Maxwell equations in vacuum,
\begin{eqnarray}
\label{Maxwell}
\rot \Ev = -\partial_t \Bv
\qquad
 \div \Bv = 0  
\cr
\rot \Bv = \partial_t \Ev
\qquad
 \div \Ev = 0  \,.
\end{eqnarray}
These equations are invariant under the duality $\Bv\to\Ev$, $\Ev\to-\Bv$.
In the following we gather the fields $\Ev$ and $\Bv$ in a 6-component vector $\Fv=\{ \Ev, \Bv \}$. In analogy with (\ref{reduction-sca}) we look for solutions of the form
\be\label{reduction}
\Fv(t,x,y,z)=e^{iPz-i\omega t}\, {\cal R}(qz)\F(X,Y) \,,
\ee
in the complex notation (the physical field is the real part of it). $\F=\{\E,\B\}$ is the field in the $z=0$ plane and ${\cal R}(qz)$ is the rotation matrix of (\ref{param}) applied to the vectors $\E$ and $\B$ :
\be\label{composantesE}
E_x=\E_X \, \cos qz-\E_Y \, \sin qz \,,\quad E_y=\E_X \, \sin qz+\E_Y\, \cos qz \quad\hbox{(idem for } \Bv).
\ee
To write the boundary conditions we introduce the vectors 
\be\label{tangent}
\vsigma(s)=\rv'(s)=\pmatrix{X'(s)\cr Y'(s)\cr 0}, 
\quad \vtau(s)=\zu+q\,\zu\times\rv(s)=\pmatrix{-qY(s)\cr qX(s)\cr 1},
\ee
which are tangent to the wall at $z=0$, and the normal vector
\be\label{normal}
\vnu(s)=\vsigma(s)\times \vtau(s)=\pmatrix{Y'(s)\cr -X'(s)\cr q(XX'+YY')}.
\ee
The boundary conditions are $\vec\E\cdot\vsigma=0$, $\ \vec\E\cdot\vtau=0$ and $	\vec\B\cdot\vnu=0$. In cylindrical coordinates, they read 
\begin{eqnarray}
\label{ET=0'}
\E_r\,\sigma_r+\E_\varphi\,\sigma_\varphi &=& 0  \,, \\
\label{ET=0''}
qr\,\E_\varphi + \E_z &=& 0  \,, \\
\label{Bn=0'}
\B_r \, \sigma_\varphi+(qr\B_z-\B_\varphi) \sigma_r  &=& 0  \,. 
\end{eqnarray}
These 3 conditions are not independent. Only two suffice. They break the duality $\Bv\to\Ev$, $\Ev\to-\Bv$.

$\Fv$ can be decomposed in a manner similar to (\ref{decompose-scalaire}). However, since the photon has two polarisation states, there are two free partial waves for each $l$, which is now the total (spin + orbital) angular momentum. 

\subsection{Study of a free partial wave}

Let us first study separately one partial wave of definite $\omega$, $l$ and $p$ (here $p=p_l=P-lq$). In cylindrical coordinates it takes the form
\be\label{OndeCyl}
\Fv_{\omega,p,l}(t,r,\varphi,z) = 
\pmatrix{E_r\cr E_\varphi\cr E_z\cr B_r\cr B_\varphi\cr B_z}= e^{-i\omega t}\, e^{ip z}\, e^{il\varphi}\,
\pmatrix{\E_r(r)\cr \E_\varphi(r)\cr \E_z(r)\cr \B_r(r)\cr \B_\varphi(r)\cr \B_z(r)}\,.
\ee
In this subsection the quantities written in caligraphic depend only on $r$. The Maxwell equations become
\begin{eqnarray}
(l/r)\,\E_z-p\,\E_\varphi &=& \omega\,\B_r\,,\cr
-\partial_r\,\E_z+ip\,\E_r &=& i\omega\,\B_\varphi\,,\cr
\left(r^{-1}+\partial_r\right)\E_\varphi - i(l/r)\,\E_r &=& i\omega\,\B_z\,,
\end{eqnarray}
and the dual equations obtained by $\B\to\E$, $\E\to-\B$. These equations are redundant and one can omit the third one and its dual. 

\paragraph{The TM/TE basis.} A first basis is made of the so-called 
\emph{transverse magnetic} (TM) and \emph{transverse electric} (TE) states. In cylindrical coordinates, 
\be\label{TM}
\F^{(\rm TM)}_{\omega,p,l}(r) =
\pmatrix{
\E_r(r)\cr \E_\varphi(r)\cr \E_z(r)\cr \B_r(r)\cr \B_\varphi(r)\cr \B_z(r)}_{(\rm TM)}
= k^{-|l|-2\delta_{l,0}}\pmatrix{
+ ip\,k\,J'_{|l|}(kr)\cr 
- p\,l/r\,J_{|l|}(kr)\cr
+ k^2\,J_{|l|}(kr)\cr 
+ \omega\,l/r\,J_{|l|}(kr)\cr
+ i\omega\,k\,J'_{|l|}(kr)\cr 
0\cr
}\,, 
\ee
with $k=(\omega^2-p^2)^{1/2}$. The TM and TE fields are dual, up to a factor $i$ : 
$\B^{(\rm TE)}=-i\E^{(\rm TM)}$,  $\E^{(\rm TE)}=i\B^{(\rm TM)}$.  
For the $l=0$ waves, 
\be\label{TMTEl=0}
\F^{(\rm TM)}_{\omega,p,0}(r) = \pmatrix{
- ip\,\psi_1\cr 
0\cr
\psi_0\cr 
0\cr
- i\omega\,\psi_1\cr 
0\cr
}\,,\qquad 
\F^{(\rm TE)}_{\omega,p,0}(r) = \pmatrix{
0\cr
+  \omega\,\psi_1\cr 
0\cr
- p\,\psi_1\cr 
0\cr
-i \psi_0\cr 
}\,,
\ee
$\psi$ being defined in (\ref{Bessel}).

\paragraph{The chiral basis.}

When $k\to0$ with $l\ne0$, $\F^{(\rm TM)}$ and $\F^{(\rm TE)}$ become equal up to a sign and they no more form a complete basis. We avoid this degeneracy with the combinations  $\F_+=\F^{(\rm TE)}+\F^{(\rm TM)}$, $\F_-=\F^{(\rm TE)}-\F^{(\rm TM)}$, which are respectively the states of right and left \emph{chirality}, or \emph{helicity} $\chi=+1$ and $\chi=-1$. They are such that $\Ev=i\chi\Bv$. 
For real $k$ a right-handed (resp. left-handed) solution is a superposition of plane waves of positive (resp. negative) helicity.  
More precisely, we normalise these states as~:
\be\label{chirale}
\F_{\omega,p,l,\chi}= -2{\F^{(\rm TE)}+\chi\F^{(\rm TM)}\over \chi p+s_l\omega}
=\pmatrix{
-i \left[ \psi_{\lambda-1} + (p-\chi'\omega)^2 \, \psi_{\lambda+1}\right] \cr
s_l\,\left[\psi_{\lambda-1} - (p-\chi'\omega)^2 \, \psi_{\lambda+1}\right] \cr
2(p-\chi'\omega) \, \psi_{\lambda} \cr
-\chi\,\left[ \psi_{\lambda-1} + (p-\chi'\omega)^2 \, \psi_{\lambda+1}\right] \cr
-i\chi'\,\left[\psi_{\lambda-1} - (p-\chi'\omega)^2 \, \psi_{\lambda+1}\right] \cr
-2i\chi\,(p-\chi'\omega) \, \psi_{\lambda} \cr
}. 
\ee
Here $s_l={\rm sign(l)}$, $\ \chi'=s_l\chi$, 
$\ \lambda=|l|$ and $\psi_{\lambda} = J_{\lambda}(kr)/k^{\lambda}$. 
Use was made of the recurrence relations 
$$
(l/x)\,J_l(x)=[J_{l-1}(x)+J_{l+1}(x)]/2\,, \qquad J'_l(x)=[J_{l-1}(x)-J_{l+1}(x)]/2\,.
$$
The denominator $(\chi p+s_l\omega)$ eliminates a kinematical zero, so that these states have finite and  linearly independent limits when $p\to-\chi'\omega$.
We will take the chiral basis except for the $l=0$ states, where we keep the TM-TE basis (\ref{TMTEl=0}). 

\subsection{The partial wave expansion}

According to (\ref{TMTEl=0}-\ref{chirale}), the decomposition (\ref{decompose-scalaire}) is replaced by
\begin{eqnarray}
\label{decompose-vectoriel}
\Fv(t,r,\varphi,z)=e^{-i\omega t}\, e^{iPz} \left[
a_0 \, \F^{(\rm TM)}_{\omega,P,0}(r) + b_0 \, \F^{(\rm TE)}_{\omega,P,0}(r) \right] \nonumber\\
+\,e^{-i\omega t}\,\sum_{l\ne0}  e^{il\varphi}\, e^{ip(l) z} \left[
a_l \, \F_{\omega,p(l),l,\chi'=+1}(r) + b_l \, \F_{\omega,p(l),l,\chi'=-1}(r) \right],
\end{eqnarray}
with $p(l)=P-lq$. For $l\ne0$ the $a_l$ (resp. $b_l$) coefficients multiply the  states where the chirality has the same sign as $l$ (resp. opposite sign to $l$). 

The boundary conditions (\ref{ET=0'}-\ref{ET=0''}) read
\begin{eqnarray}\label{ET=0-3}
&0& = (p\sigma_r\, a_0 + i\omega\sigma_\varphi\, b_0)\,\psi_1 \nonumber\\	
+ \sum_{l\ne0}e^{il\varphi} &\{& 
a_l \left[ \,(\sigma_r+is_l\sigma_\varphi) \, \psi_{\lambda-1} 
+(p-\omega)^2\,(\sigma_r-is_l\sigma_\varphi)\,\psi_{\lambda+1} \right]\nonumber\\
&+& b_l \left[ \,(\sigma_r+is_l\sigma_\varphi) \, \psi_{\lambda-1}
+(p+\omega)^2\,(\sigma_r-is_l\sigma_\varphi) \,\psi_{\lambda+1} \right]
\,\}\,,  
\end{eqnarray}
\begin{eqnarray}
\label{ET=0-4}
&0&= a_0\,\psi_0 + \omega qr\,\psi_1 \, b_0 \nonumber\\	
+ \sum_{l\ne0}e^{il\varphi} &\{& 
a_l \left[s_l\,qr\,(\psi_{\lambda-1}-(p-\omega)^2\,\psi_{\lambda+1}) 
+2(p-\omega)\,\psi_{\lambda}\right]
\nonumber\\
&+& b_l \left[s_l\,qr\,(\psi_{\lambda-1}-(p+\omega)^2\,\psi_{\lambda+1})
+2(p+\omega)\,\psi_{\lambda}\right]
\,\}\,. 
\end{eqnarray}
In these equations, $\lambda=|l|$, $\ p=p_l=P-lq$,  
$\ k=k_l=\left[\omega^2-p_l^2\right]^{1/2}$, the argument of $\psi_\lambda$, $\psi_{\lambda-1}$, and $\psi_{\lambda+1}$ is $k_lr(s)$, $\ \varphi=\varphi(s)$ and 
$\vsigma = \vsigma(s)$ is given by (\ref{tangent}).

Generalising the truncation-and-sutture method of Section 1.2 we write conditions (\ref{ET=0-3}-\ref{ET=0-4}) for $N$ points $\rv_n$ of polar coordinates $(r_n,\varphi_n)$, keeping only $N$ values of $l$. It gives $2N$ linear equations linking $N$ coefficients $a_l$ and $N$ coefficients $b_l$. The dispersion relation is the vanishing of the $2N\times2N$ determinant of this system, the coefficients of which are given in Appendix (\ref{coef2}-\ref{coef20}). $\mathcal{L}$ and $\rv_1,...\rv_N$ are chosen like in the scalar case.

We have tested this method with a straight ($q=0$) cylindrical wave guide. The well-known TM and TE lower modes were found using no more than ten points, even when the $Oz$ axis was not the guide axis, but shifted by about half a radius.

\section{Application to the acceleration of relativistic particles}

A particle moving along the axis of the helical wave guide is accelerated or decelerated by the $E_z$ component at $r=0$, whose physical value is $\Re \{a_0 e^{iPz-i\omega t}\}$. Besides, the $l=\pm1$ partial waves give a transverse field at $r=0$, the components of which are, in the rotating frame 
\begin{eqnarray}
E_X&=&\Im \{(a_1 + b_1 + a_{-1} + b_{-1}) \, e^{iPz-i\omega t}\} \cr
E_Y&=& \Re \{(a_1 + b_1 - a_{-1} - b_{-1}) \, e^{iPz-i\omega t}\} \cr
B_X&=&\Re \{(- a_1 + b_1 + a_{-1} - b_{-1}) \, e^{iPz-i\omega t}\} \cr
B_Y&=&\Im \{(a_1 - b_1 + a_{-1} - b_{-1}) \, e^{iPz-i\omega t}\}\,. 
\end{eqnarray}
If an electron moves along the axis at the same velocity as the wave ($z=v(t-t_0)$ with $v=v_{\rm ph}$), it will experience a Lorentz force ${\bf f}= (-e)\,(\Ev+\vv\times\Bv)$ which is constant in the rotating frame. In the ultrarelativistic case ($v_{\rm ph}\simeq1$), this force is 
\be
\pmatrix{f_X\cr f_Y} \simeq (-e)\times 2\Re \left\{e^{-i\omega t_0}\left[
b_1\pmatrix{-i\cr1} -  b_{-1}\pmatrix{i\cr1}
\right]\right\} \,.
\ee
In the laboratory frame, it is the force excerted by a helical undulator having the same period as the helix. We therefore get a parasitic undulator radiation. 
It can be avoided with a base which has a symmetry of order $S\ge2$ about $Oz$, choosing a mode involving only the waves $l=0$ modulo $S$.

\section{Numerical results}

Like in the scalar case (Section \ref{numeric-scal}) we took a circular base with parametres $\varepsilon=0.4$, $q=1$ and imposed the phase velocity $v_{\rm ph}= 1-10^{-3}$. Plotting $\ln|\det\{C_{nl}\}|$ or, better, $\arg|\det\{C_{nl}\}|$ 
we located modes at $\omega a=2.28$, $3.6$, $4.8$ and $5.0$ which are stable at least at the $0.1$ precision for $N\sim15$ (they were the same for $\mathcal{L}=[-6,+8]$, $[-5,+9]$ and $[-3,+11]$). The first three modes seem to be mostly of the TE type~: $|E_z(0)/B_z(0)|$ = 0.3, 0.25 and 0.4. We give below more details about the fourth one ($\omega a=5.0$) which is mainly TM~: $|E_z(0)/B_z(0)| = 5.4$.

Fig.3 
shows the cross section of the fields along the segment $x\in[-0.6,+1.4]$, $y=0$, which is a diametre of the base circle (in units  $a=1$). At the extremities the boundary condition $B_r=0$, which was not directly imposed by (\ref{ET=0-3}-\ref{ET=0-4}) is well satisfied. 

\begin{figure}
\centering{%
\begin{tabular}{c c}
\includegraphics[width=.49\textwidth]{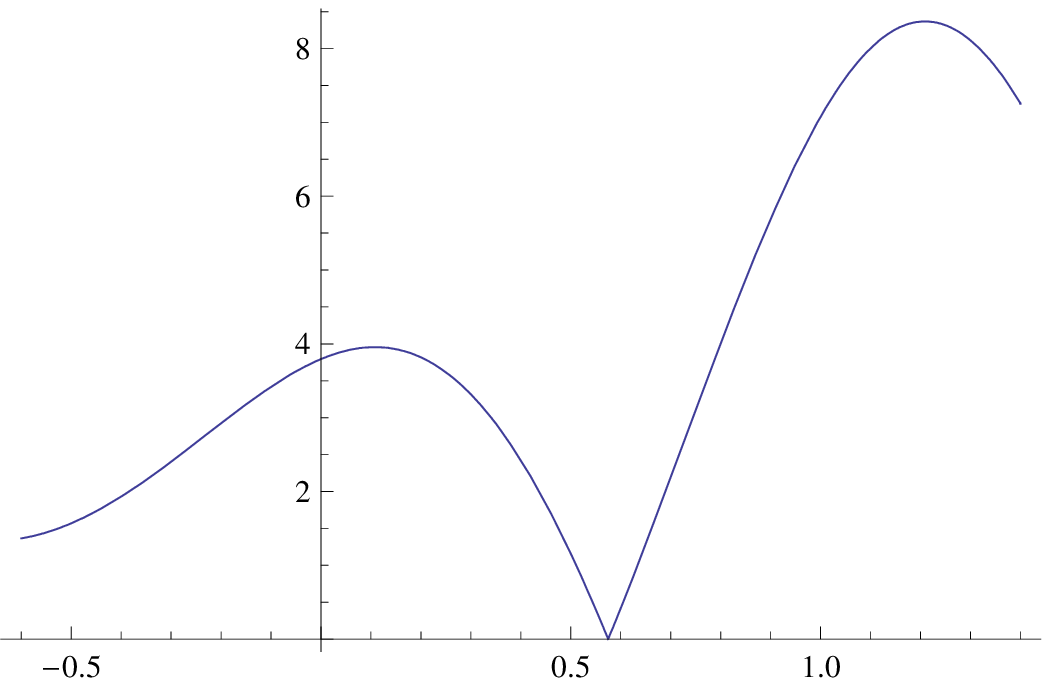} 
&
\includegraphics[width=.49\textwidth]{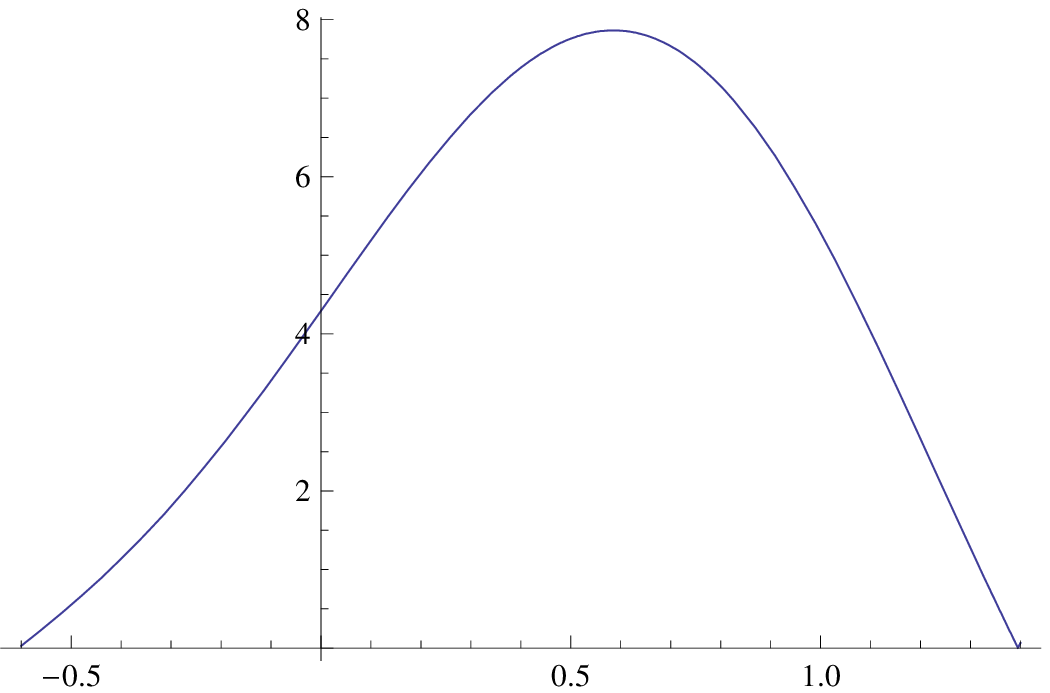} 
\\
\includegraphics[width=.49\textwidth,]{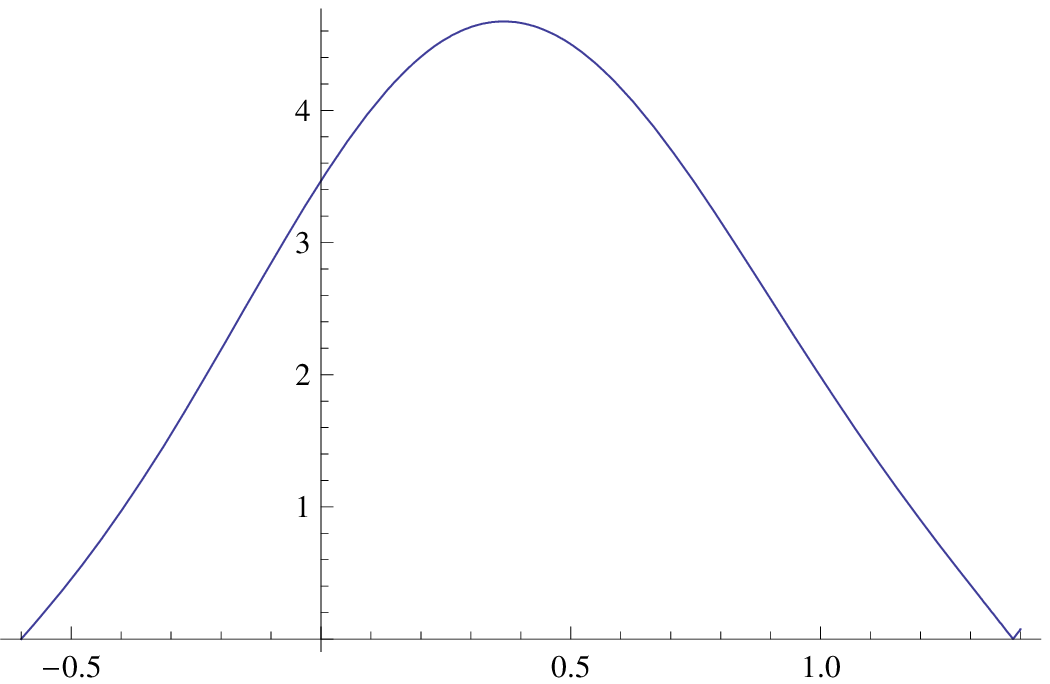} 
&
\includegraphics[width=.49\textwidth,]{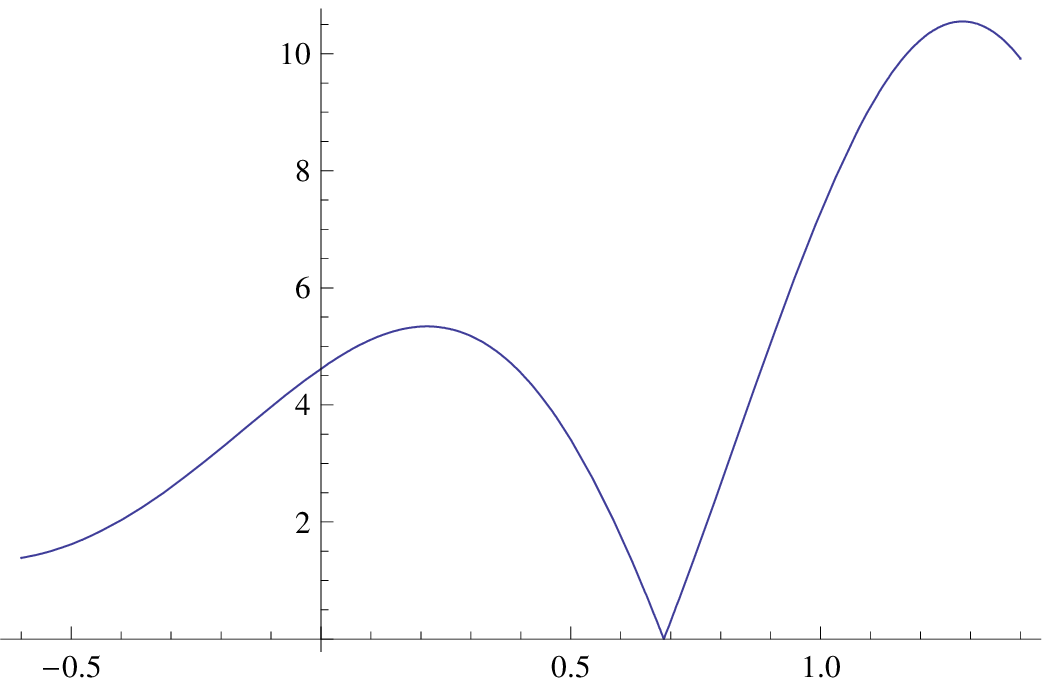} 
\\
\includegraphics[width=.49\textwidth]{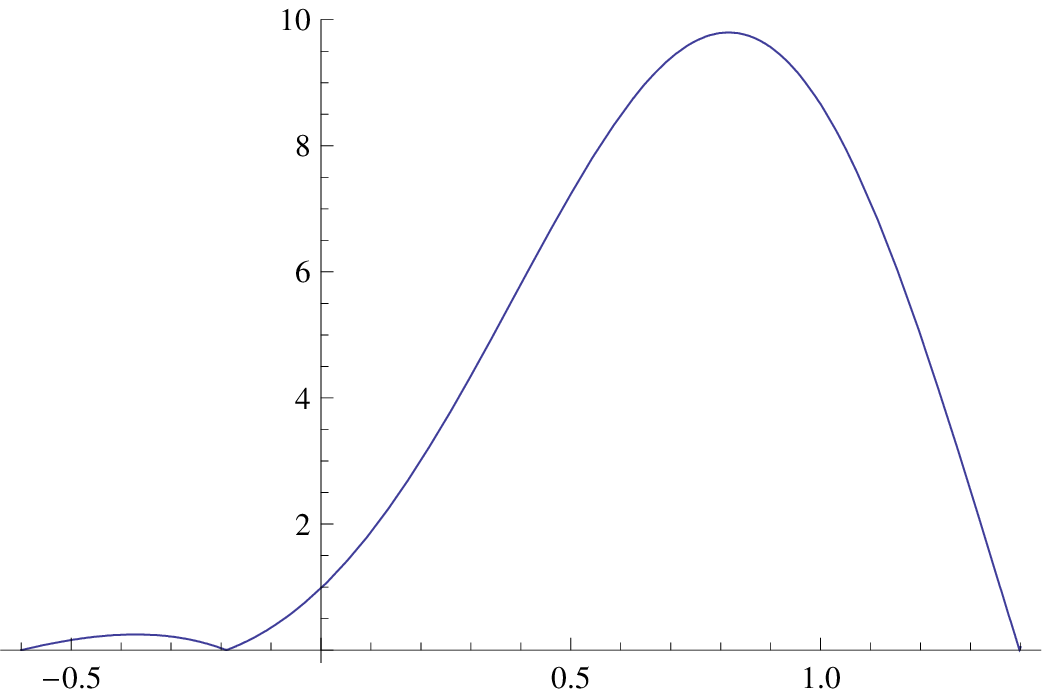} 
&
\includegraphics[width=.49\textwidth]{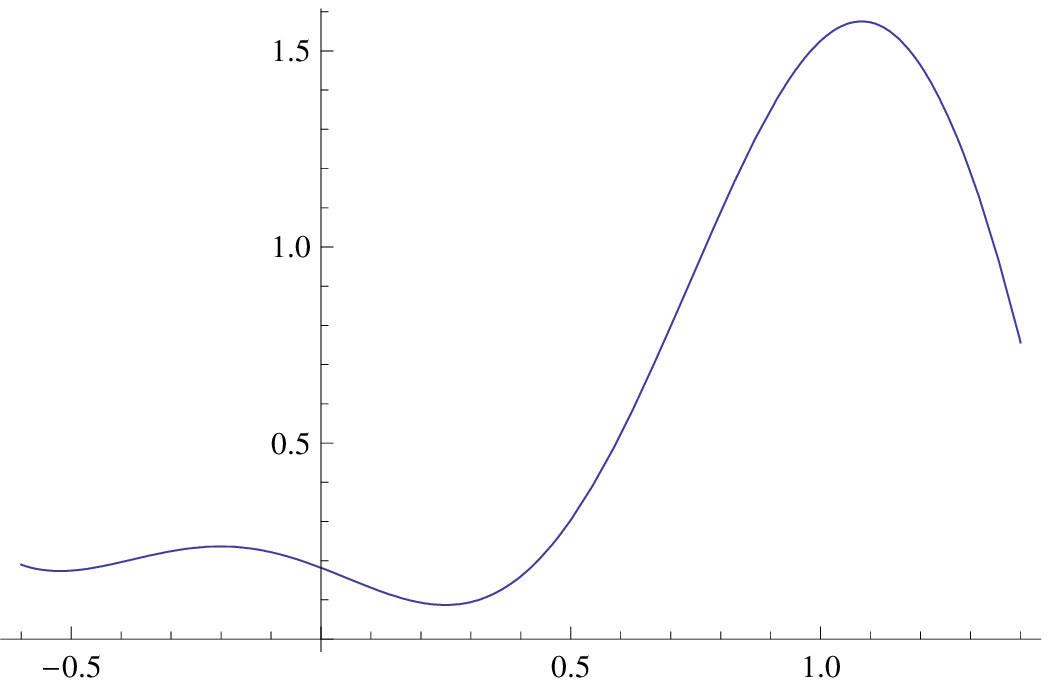}
\end{tabular}
}
\caption{Cross section of the electromagnetic field $\F(x,y)=\Fv(0,x,y,0)$ along the line $y=0$ for the TM-like mode at $\omega a=5.0$ (circular base, $\varepsilon=0.4$ and $q=1$). 
Left curves, from top to bottom : $|\E_r(x,y)|$,  $|\E_\varphi(x,y)|$ and $|\E_z(x,y)|$ at $y=0$.
Right curves~: same for $\B_r$,  $\B_\phi$ and $\B_z$.}
\end{figure}

\begin{figure}\label{3D} 
\centering{%
\begin{tabular}{c c}
\includegraphics[width=.49\textwidth]{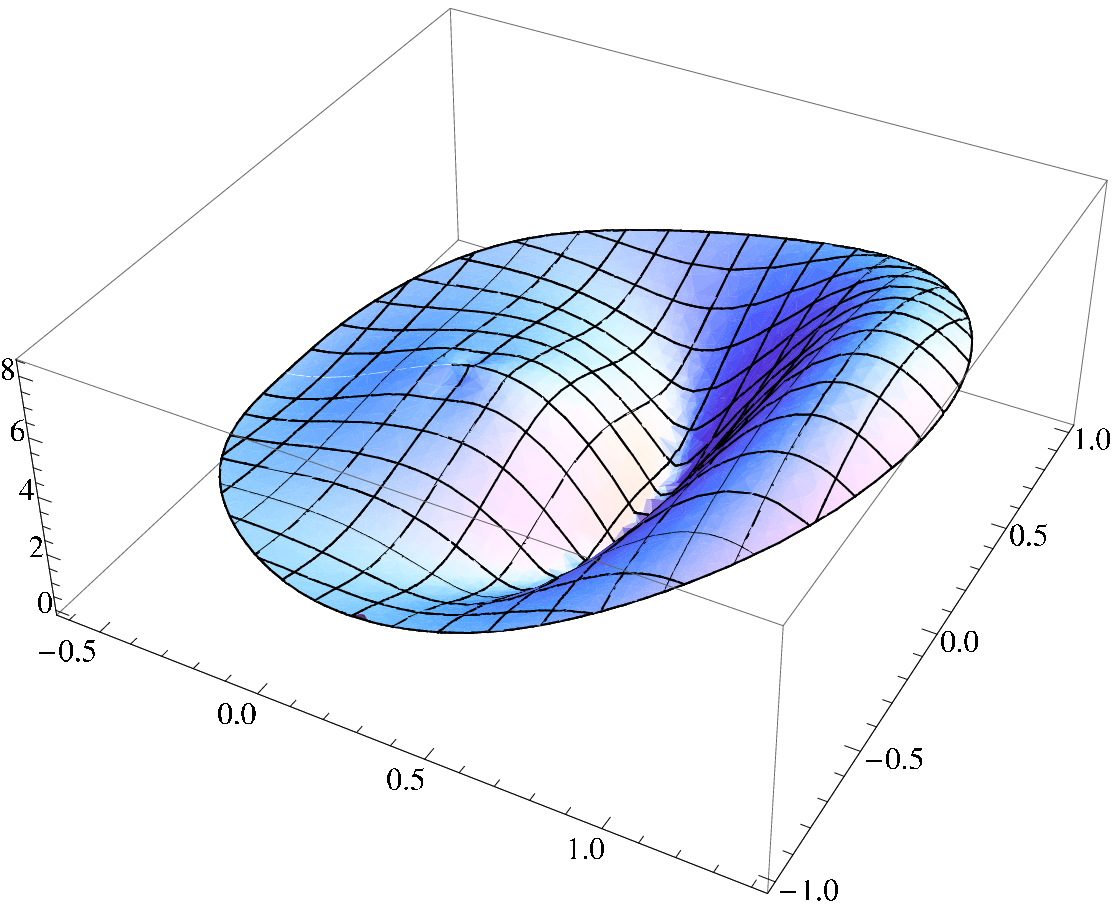}  
&
\includegraphics[width=.49\textwidth]{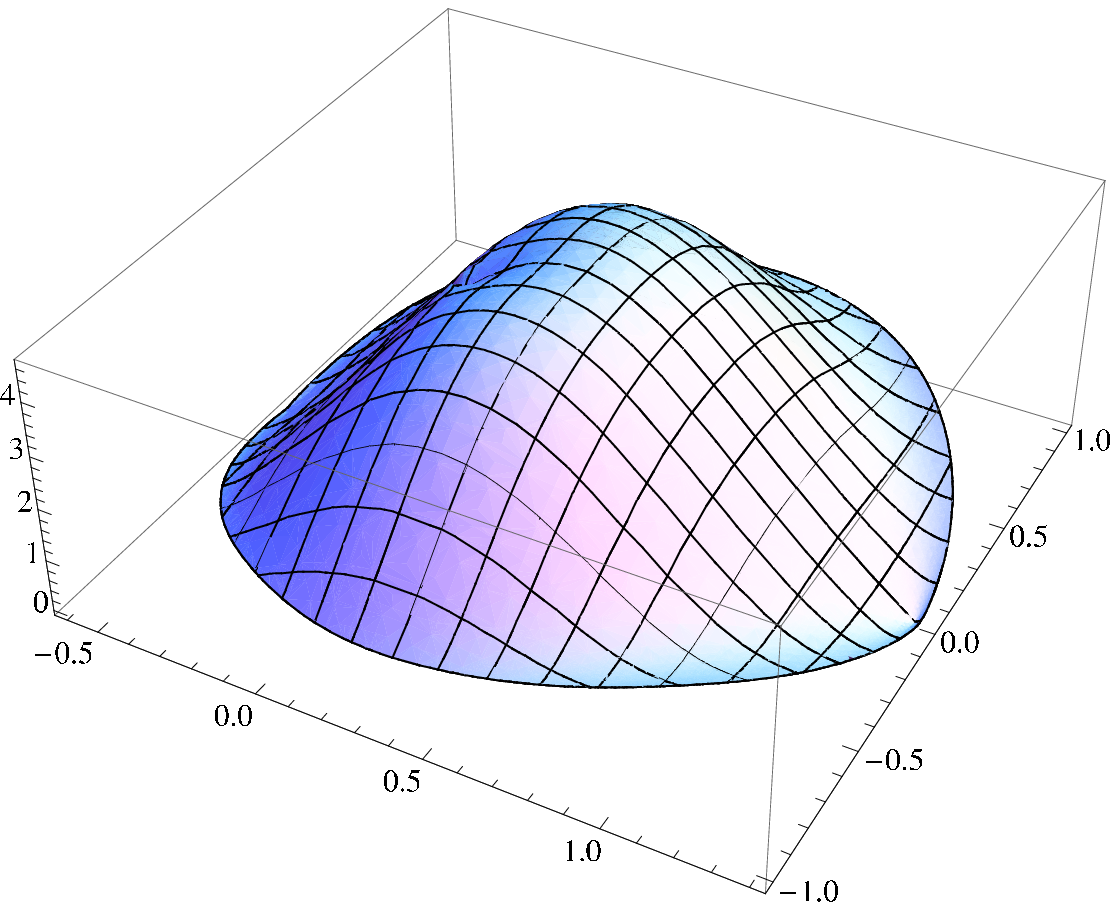} 
\\
\includegraphics[width=.49\textwidth,]{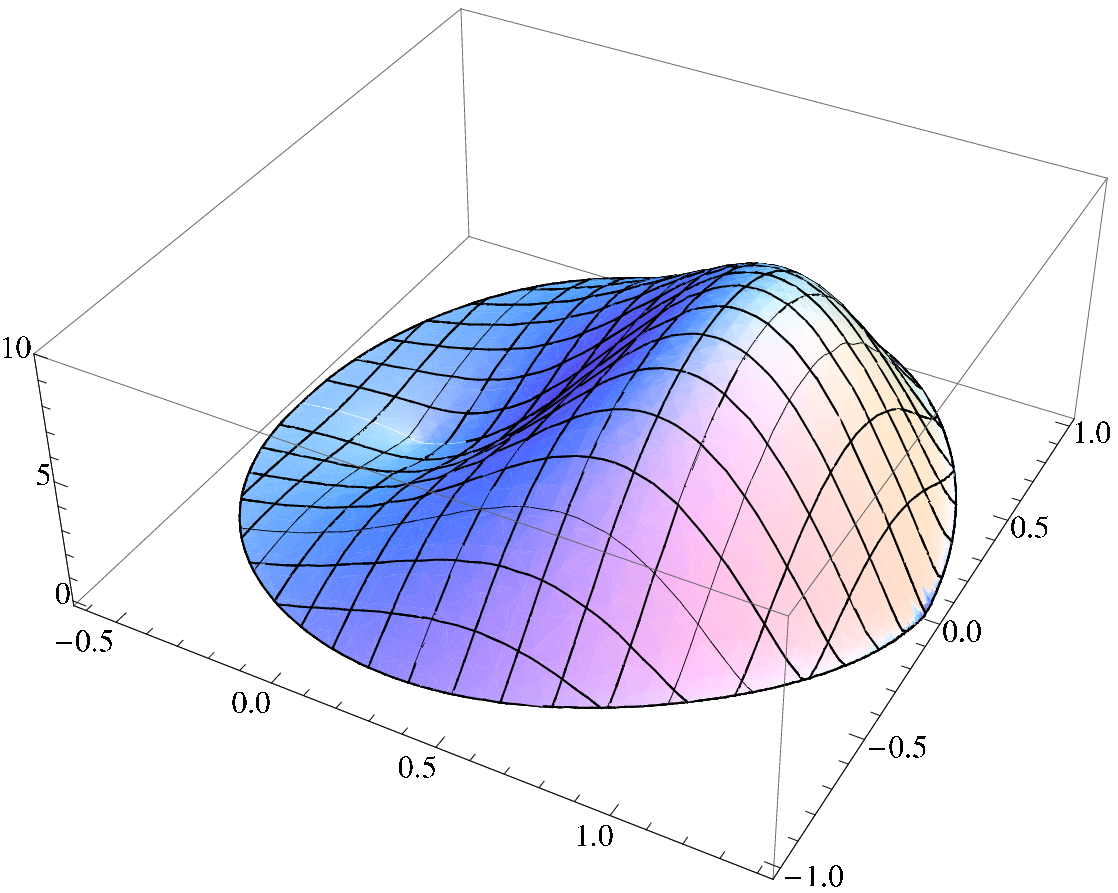} 
&
\includegraphics[width=.49\textwidth,]{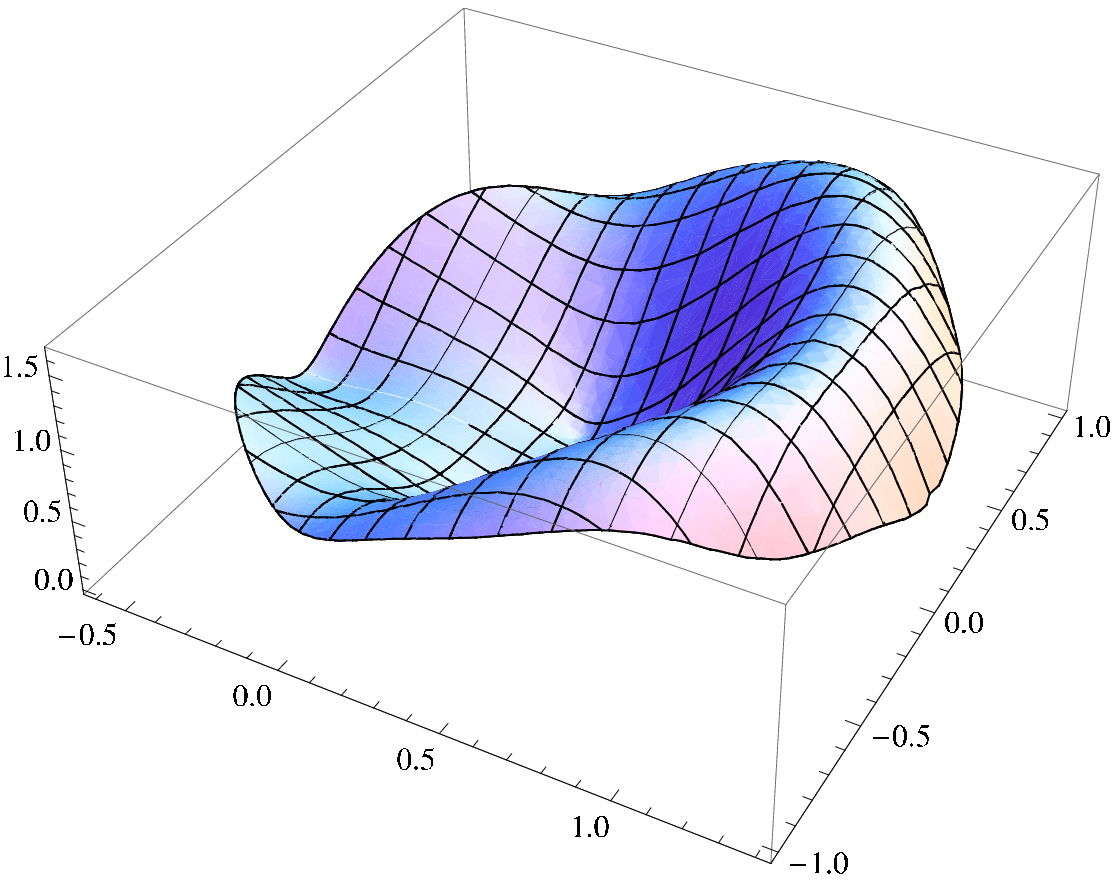} 
\end{tabular}
}
\caption{3-dimensional maps of $|E_r|$, $|E_\varphi|$ (upper mapps), $|E_z|$ and $|B_z|$ (mower mapps) for the same mode as Fig.3. 
}
\end{figure}

Fig.\ref{3D} shows the 3-dimensional plot of $|\E_r(x,y)|$, $|\E_\varphi(x,y)|$, $|\E_z(x,y)|$ and $|\B_z(x,y)|$. We did not plot $|\B_r|$ and $|\B_\varphi|$, which look like $|\E_\varphi|$ and $|\E_r|$.

A \emph{figure of merit} of the helical wave tube as an accelerating device is the ratio of the accelerating field over the strongest electric field at the boundary, 
\be
f.o.m. = E_z(0)/\max(|E(\rv(s))|)\,. 
\ee 
Indeed a too strong field at the boundary may trigger an electron emission. Fig. \ref{3D} indicates that the strongest boundary field is at $r=r_{\rm max}$ and Fig.3 
gives $f.o.m.\simeq1/8$.

Another relevant quantity is the group velocity $v_{\rm g}=d\omega/dP$, which can be easily calculated from (\ref{det-scal}) when the phase velocity $v_{\rm ph}=\omega/P$ is not imposed. For the $\omega a=5.0$ mode studied above, $v_{\rm g}=0.68$. The guide can therefore be quickly filled with the wave and quickly emptied. To save energy, one may use a recirculated wave.

We also have preliminary results for a waveguide with a stronger helical bending: $\varepsilon=0.6$, $q=1.5$. Imposing $v_{\rm ph}= 1-10^{-3}$, three first modes are seen at $\omega a=2.2$, $3.3$ and $4.0$ with $|E_z(0)/B_z(0)|$ = 0.35, 0.3 and 0.8 respectively. The number of sutture points, $N\sim 10-20$, was enough to obtain  stable roots but not yet stable field amplitudes. This is like in the variational method in quantum mechanics, where the energy converges faster than the wave function. 

\newpage
\section{Conclusion}

This study has shown that a linear accelerator for ultrarelativistic particles may work with a helical wave guide. Until now we have made detailed numerical calculations only for a guide with circular base, one set of parametres $\{\varepsilon,q\}$ and the lower TM-like mode. The figure of merit  (\emph{accelerating field}) / (\emph{strongest electric field at the boundary}) of this mode was only about 1/8, but we hope that different guide parametres or different base shapes can give higher values. 

The study has also validated the truncation-and-sutture method. This economical method seems to be well suited to calculate the modes of a scalar or vector field in a 2- or 3-dimensional homogeneous cavity of arbitrary, but smooth enough shape.


\section{APPENDIX : The $M$ matrix}
Defining

\ni$~C_{n,l}$ = coefficient of $a_l$ in the condition $\vec\E\cdot\vsigma=0$ at $\rv_n$,

\ni$~C_{n,l+N}$ = coefficient of $b_l\equiv a_{l+N}$ in the condition $\vec\E\cdot\vsigma=0$ au point $\rv_n$,

\ni$~C_{n+N,l}$ = coefficient of $a_l$ in the condition $\vec\E\cdot\vtau=0$ at $\rv_n$,

\ni$~C_{n+N,l+N}$ = coefficient of $b_l$ in the condition $\vec\E\cdot\vtau=0$ at $\rv_n$,

\ni the $2N\times2N$ matrix of the linear system of equations expressing the boundary conditions is decomposed in four $N\times N$ submatrices~:

\be\label{coef1}
	M=\pmatrix{	...& ... & ... & | & ...& ... & ...\cr
	...& C_{n,l} & ... & | & ...& C_{n,l+N} & ...\cr
	...& ... & ... & | & ...& ... & ...\cr
		-& - & - & | & - & - & -\cr
		...& ... & ... & | & ...& ... & ...\cr
	...& C_{n+N,l} & ... & | & ...& C_{n+N,l+N} & ...\cr
		...& ... & ... & | & ...& ... & ...	}
\ee
The matrix elements are, for $l\ne0$,
\begin{eqnarray}\label{coef2}
	C_{n,l} &=& \left[ \psi_{\lambda-1}\,\hat\sigma^{s_l}
+(p-\omega)^2\,\psi_{\lambda+1}\,\hat\sigma^{-s_l} \right] \, e^{il\varphi_n} \cr
	C_{n,l+N} &=& \left[ \psi_{\lambda-1}\,\hat\sigma^{s_l}
+(p+\omega)^2\,\psi_{\lambda+1}\,\hat\sigma^{-s_l} \right] \, e^{il\varphi_n} \cr
	C_{n+N,l} &=& \left[s_l\,qr\,(\psi_{\lambda-1}-(p-\omega)^2\,\psi_{\lambda+1}) 
+2(p-\omega)\,\psi_{\lambda}\right] \, e^{il\varphi_n} \cr
	C_{n+N,l+N} &=& \left[s_l\,qr\,(\psi_{\lambda-1}-(p+\omega)^2\,\psi_{\lambda+1})
+2(p+\omega)\,\psi_{\lambda}\right] \, e^{il\varphi_n} 
\,,
\end{eqnarray}
and for $l=0$,
\begin{eqnarray}\label{coef20}
	C_{n,0} &=& P\,\psi_1\,\sigma_r \cr
	C_{n,N} &=& i\omega \,\psi_1\,\sigma_\varphi	\cr
	C_{n+N,0} &=& \psi_0	\cr
	C_{n+N,N} &=& \omega q r\,\psi_1\,.
\end{eqnarray}
In these formulas
$p=p_l=P-lq$, $\ k=k_l=\left[\omega^2-(P-lq)^2\right]^{1/2}$, $\lambda =|l|$, $\ s_l={\rm sign(l)}$, $\ \psi_\lambda = J_\lambda(k_l r_n)/k_l^\lambda$ and 
we have introduced the unitary complex number
$$
\hat\sigma_n={\sigma_r(\rv_n)+i\sigma_\varphi(\rv_n)\over |\vec\sigma(\rv_n)|}
=e^{-i\varphi_n}\,
{\sigma_x(\rv_n)+i\sigma_y(\rv_n)\over |\vec\sigma(\rv_n)|}
\,.
$$

\end{document}